\documentclass[12pt,a4paper]{article}
\usepackage{graphicx}
\usepackage{amsmath}

\begin{document}

\vspace*{-1cm}

\begin{center}
{\large \textbf{Exact periodic solutions of the Liouville equation \\[0pt]
and the \ ``snake'' of density in JET}}

\smallskip 

\underline{F. Spineanu} and M.Vlad

\textit{Association Euratom-MEC Romania \\[0pt]
NILPRP, P.O.Box MG-36, Magurele, Bucharest, Romania}
\end{center}


The \emph{snake} phenomenon has been observed in JET during the experiments
of pellet injection and consists in formation of persistent density
perturbations at rational-$q$ surfaces. These structures persist over
several sawtooth collapses and are difficult to explain as magnetic
perturbations. Possibly related to this, there are indications that the
tokamak plasma density has an anomalous radial pinch, much larger than that
of the neoclassical origin. In this class of phenomena one should also
include the persistent impurity accumulation in laser blow-off injected
impurity, observed in experiments in TCV. There are several studies of the
statistical properties of the correlations between the peaking factors for
density, current density or pressure, with plasma parameters and these
studies seem to support the idea of turbulent equipartition of the
theromodynamic invariants (Minardi). However we should note that these
studies involve quantities expressed as global variables (like averages) and
they can hide other dependences not immediately obvious.

We consider the possibility that the particle density behavior (and
particularly the \emph{snake} phenomenon) can be connected with the
existence of attracting solutions of certain nonlinear integrable equations.
The statistical studies carried out on a large set of discharges with the
purpose of testing the prediction of the Turbulent Equipartition theory have
suggested that the current density is given by the equation $\Delta j+\left(
\lambda ^{2}/4\right) j=0$ where $\lambda $ is a constant. However this may
be valid on finite spatial \emph{patches} and on every patch the equation
can be considered an approximation of some more general equation, for which
the derivation from first principles may be possible. One is the \emph{sinh}%
-Poisson equation, $\Delta \phi +\left( \lambda ^{2}/4\right) \sinh \phi =0$
for which the previous equation is the linearised form. The reason to
consider the \emph{sinh}-Poisson equation comes from the existing proofs
that this equation governs the asymptotic states of ideal fluids, or, more
generally, of $2D$ systems that can be reduced to the dynamics of point-like
elements interacting by the potential which is the inverse of the Laplacean
operator (Jackiw and Pi, Spineanu and Vlad). This equation is however
obtained when there are two kinds of elements (like positive and negative
vorticity) and they are of equal number, $n_{+}=n_{-}$. Then the \emph{sinh}%
-Poisson equation is obtained as governing the states with maximum entropy
of the discrete statistical system at negative temperature (Montgomery \emph{%
et al.}). Since the equation for the current density mentioned above is
derived under the assumption of turbulent equipartition, the two
descriptions may be related. However, the solutions for the \emph{unbalanced}
system of elements, $\alpha \equiv n_{+}/n_{-}\neq 1$, $\Delta \phi =\left(
\lambda ^{2}/8\right) \left[ \exp \left( \phi \right) /\sqrt{\alpha }-\sqrt{%
\alpha }\exp \left( -\phi \right) \right] $ have been obtained numerically
(Pointin and Lundgren) and have been shown to have higher entropy and higher
stability than those of the \emph{sinh}-Poisson, precisely the
characteristics we are seeking for. The limiting form of the unbalanced
equation is the Liouville equation, $\Delta \phi =\left( \lambda
^{2}/8\right) \exp \left( -\phi \right) $.

There are field theoretical models that are able to describe statistical
ensembles of a discrete sets of elements and these models lead to the
Liouville equation when a particular condition (called : \emph{self-duality}%
) is fulfilled. We note that the Liouville equation can lead to the above
current density equation (Helmholtz-type) in an approximation where the
function differs weakly from a background value.

\bigskip

In the following we will discuss solutions of the soliton type of the
Liouville equation $\Delta \phi =\left( \lambda ^{2}/8\right) \exp \left(
-\phi \right) $. It is commonly considered that the Liouville equation is
solved by the formula given in terms of two arbitrary complex functions $F$
and $G$ 
\begin{equation*}
\phi \left( x,y\right) =-\ln \left\{ \frac{F^{\prime }\left( z\right)
G^{\prime }\left( z^{\ast }\right) }{\left[ 1+\frac{\lambda ^{2}}{16}F\left(
z\right) G\left( z^{\ast }\right) \right] ^{2}}\right\}
\end{equation*}
where $z=x+iy$ and $^{\ast }$ is the complex conjugate. However this form of
the solution is too general to be useful. More popular is the particular
form derived from this one, 
\begin{equation*}
\phi =\ln \left[ \cosh \left( kx\right) +\varepsilon \cos \left( ky\right) %
\right]
\end{equation*}
solution of $\Delta \phi =\left( 1-\varepsilon ^{2}\right) k^{2}\exp \left(
-2\phi \right) $. This describes the \emph{cat's eye} vortex chains of the
Kelvin-Stuart stream function, and the magnetic flux function in the chains
of magnetic islands (Finn and Kaw).

Our main objective is to identify a type of solution of the Liouville
equation which is more localised on a magnetic surface compared with the
profile of a magnetic island of the same helical symmetry. For this the
factor $k$ must be independent of the local $q$: the spatial extension of
the density perturbation at a magnetic surface and the helical symmetry of
that magnetic surface must be independent. This is clear in the case of the
Liouville (and \emph{sinh}-Poisson) equation, since it is \emph{conformally}
invariant. We cannot be sure that the solution we are looking for can be of
the form shown above. The correct approach is to find any solution of the
Liouville equation.

We now describe the systematic way of obtaining solutions of the Liouville
equation on periodic domains, taking $y$ as the poloidal and $x$ as the
radial coordinate. The method (Tracy \emph{et al.}) consists of taking the
Liouville equation as the limit of the \emph{sinh}-Poisson equation: 
\begin{equation*}
\lambda ^{2}=\widetilde{\lambda }^{2}\exp \left( -\beta \right) ,\;\psi
\equiv \phi -\beta 
\end{equation*}
and let $\beta \rightarrow \infty $. Then the \emph{sinh}-Poisson equation
becomes the Liouville equation after taking $\widetilde{\lambda }\rightarrow
\lambda $. The \emph{sinh}-Poisson equation is exactly integrable on
periodic domains since it possesses a pair of Lax operators (Ting, Chen and
Lee). The eigenvalue problem for the Lax operators identifies a spectrum of
complex values where the two Bloch functions are not independent (for one
eigenvalue we have only one eigenfunction). These nondegenerate eigenvalues
are called \emph{main spectrum} and represent branching points in the
two-sheeted Riemann surface associated to the Wronskian of the two periodic
solutions. It is shown that the problem of finding the unknown function $%
\phi $ is mapped on the problem of motion of a set of functions (\emph{%
auxiliary spectrum}) on this Riemann surface. The equations of motion are
nonlinear but they can be solved exactly, using the Abel transform. In this
transformation the two-sheeted Riemann surface is mapped onto a compact
hyperelliptic Riemann surface and the equations of motion are linearised as
rotations along the cycles on this complex curve. The topology of the
hyperelliptic Riemann surface is that of a sphere with a number of handles
(the \emph{genus} of the surface) which is given by the geometry of the cuts
in the complex plane needed to uniformize the first Riemann surface.
Therefore the \emph{genus} is actually determined by the \emph{main spectrum}
of the Lax linear operator, or, in concrete terms, by the boundary condition
we require for the solution. The linear equations of motion are integrated
leading to \emph{phases}, which are linear combinations of the original
variables, $x$ and $y$. The number of phases is the number of type $A$
cycles on the surface (like the short turn on a torus), or, the \emph{genus}
; so, the number of phases is again determined by the \emph{main spectrum}.
We have to return to the original framework, and this represents the Jacobi
inversion problem. It is solved exactly in terms of Riemann \emph{theta}
functions. The exact solution of the \emph{sinh}-Poisson equation in terms
of Riemann \emph{theta} function, $\Theta $, is 
\begin{equation*}
\phi \left( x,y\right) =2\ln \left( \frac{\Theta \left( \mathbf{l}+\frac{1}{2%
}\mathbf{1}\right) }{\Theta \left( \mathbf{l}\right) }\right) 
\end{equation*}
where $\mathbf{l}=\mathbf{k}_{x}x+\mathbf{k}_{y}y+\mathbf{l}_{0}$ , $\mathbf{%
l}_{0}$ is a vector of constants, initial phases, and 
\begin{eqnarray*}
k_{x,j} &\equiv &\left( -1\right) ^{N}\frac{C_{jN}}{8\sqrt{Q}}+2C_{j1} \\
k_{y,j} &\equiv &i\left( -1\right) ^{N}\frac{C_{jN}}{8\sqrt{Q}}-2iC_{j1}
\end{eqnarray*}
The physical content of the problem is in the square matrix $C$ whose
dimension is half the number of eigenvalues in the main spectrum. The matrix 
$C$ is obtained from integrals of a basis of differential one-forms defined
on the hyperelliptic Riemann surface along the basis of closed paths
(cycles). These integrals can be converted into integrals along closed paths
on the plane of the spectral variable, around cuts or crossing these cuts.
The geometrical aspect of this conversion is numerically complicated due to
the jumps of the phases of the complex integrand at crossing the cuts.
However, the symmetries of the main spectrum allows us to use general forms
of the matrix 
\begin{eqnarray*}
C_{ij} &=&16^{N-2j+1}C_{i,N-j+1}^{\ast },\;j\leq N/2 \\
C_{ij} &=&C_{N-i+1,j}^{\ast }
\end{eqnarray*}
\begin{figure}[tbph]
\centerline{\includegraphics[height=9cm]{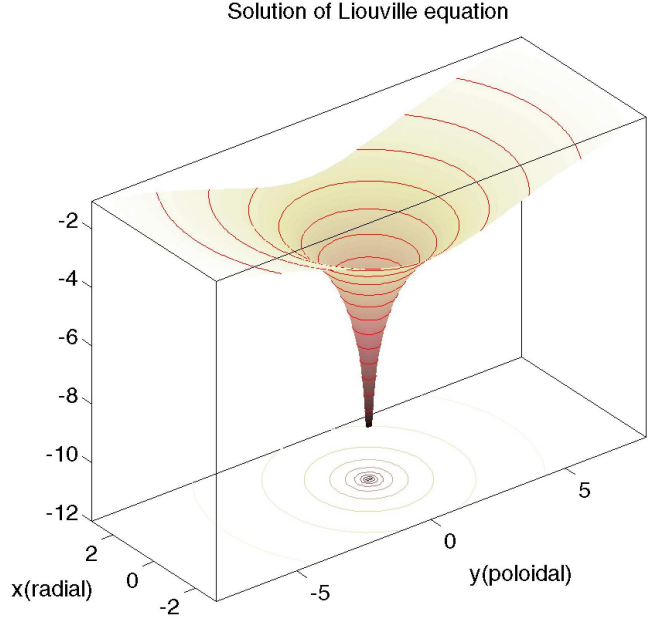}}
\caption{The streamfunction $\protect\phi $ solution of the Liouville
equation.}
\end{figure}
A particular choice of the entries of $C$ (which obeys the symmetries)
corresponds, physically, to a particular form of the boundary condition
assumed for $\phi $, on a linear section of the periodic domain.

\bigskip

The solution of the Liouville equation can be obtained from that for the 
\emph{sinh}-Poisson equation in a certain limit. This limit has been
translated into a particular distribution of the functions of the auxiliary
spectrum (Tracy \emph{et al.}). For any $\left( x,y\right) $ there are three
classes according to the positions relative to the inversion circle, which
is given by $\left| E\right| ^{2}=\lambda ^{4}/256$, $E$ being the spectral
variable. First, one notes that the discrete points of the \emph{main
spectrum} are situated in certain positions around this circle: (1) there
are $N$ inversion pairs, $\left( E_{j},E_{N+j}\right) $, in the set $E_{1}$, 
$E_{2}$, $...$, $E_{2N}$ with: $E_{N+j}=\lambda ^{4}/\left( 256E_{j}^{\ast
}\right) $ for $j=1,N$. (2) there are $M$ pairs $E_{2N+1}$, $...$, $%
E_{2N+2M} $ such that $E_{2N+k}=\lambda ^{2}\alpha _{k}/16$, $%
E_{2N+M+k}=\left( \lambda ^{2}/16\right) /\alpha _{k}^{\ast }$, with $\alpha
_{k}$ independent of $\lambda $. From each pair of the eigenvalues of the
set (2) (situated near the inversion circle) there are $M$ auxiliary
functions, which scale as $\lambda ^{2}$. The rest of the points of the
auxiliary spectrum are devided into two classes. The first contains the
auxiliary functions which are outside the inversion circle and are
independent of $\lambda $. The second are defined inside the inversion
circle and are scaled as $\lambda ^{4}$. In this way, $\lambda ^{2}\exp
\left( -\phi \right) $ is independent of $\lambda $. This classifications
cannot be directly employed, but they suggest a particular choice for the
points of the main spectrum. These studies, mainly numerical, are still in
progress.

\bigskip

A conclusion can be drawn at this stage: the solution exhibits a localised
perturbation on the poloidal direction, while the helical symmetry is still
that given of the $q$ of the surface. This solution is solitonic and
therefore is stable (we still have to clarify the effect of the limiting
procedure on the set of invariants when passing from \emph{sinh}-Poisson to
Liouville). The fact that this kind of solution is an attractor comes from
the general property of the plasma state: the self-duality (leading to the
Liouville equation) corresponds to the extremum of the action of the system
of discrete elements.

\end{document}